\documentclass{ws}

\begin{document}

\title{Numerical modelling of correlations}

\author{O.V.Utyuzh and G.Wilk}

\address{The Andrzej So\l tan Institute for Nuclear Studies; 
Ho\.za 69; 00-689 Warsaw, Poland\\E-mail: utyuzh@fuw.edu.pl and
wilk@fuw.edu.pl}

\author{Z.W\L odarczyk}

\address{Institute of Physics, \'Swi\c{e}tokrzyska Academy;
Konopnickiej 15; 25-405 Kielce, Poland\\
E-mail: wlod@pu.kielce.pl}


\maketitle

\abstracts{
We demonstrate algorithm for numerical modelling of Bose-Einstein
correlations (BEC) formulated on quantum statistical level for
a single event and exploring the property that identical particles
subjected to Bose statistics do bunch themselves in a maximal
possible way in the same cells in phase-space.} 

Bose-Einstein correlations (BEC) between identical bosons are
supposed to provide information on space-time development of 
multiparticle production processes \cite{BEC}. The usual Monte Carlo
event generators modelling such processes \cite{GEN}, because their
probabilistic structure, excludes {\it a priori} the genuine BEC,
which are of purely quantum statistical origin. One can only {\it
model} BEC by $(i)$ suitably changing output of these generators
\cite{LS,WE,AFTER} or by $(ii)$ building generator, which properly
incorporates the bosonic character of produced particles \cite{OMT}.
In both cases the goal is to reproduce the experimental two-particle
correlation function $C_2(Q=|p_i - p_j| ) = N_2(p_i,p_j)/N_1(p_i)
N_1(p_j)$. In $(i)$ this is achieved by suitable bunching of the
finally produced identical particles in phase-space performed using
special weights constructed from the output of the event generator.
In $(ii)$ the particles are already being produced in properly
bunched way by means of special generator constructed using specific
statistical model (providing Bose-Einstein or geometrical
distribution of particles in each bunch which is identified with a 
single emitting cell in phase space \cite{OMT}). Whereas $(i)$ can be
applied only for all events and is (via weights) sensitive to the
space-time structure of the production process provided by event
generator, the $(ii)$ applies already on a single event level but its
generator bears no a priori information on the space-time structure
of the production process, it uses instead nonstatistical character
of fluctuations it produces.

We propose generalization of the second approach to make it
applicable also to other generators\footnote{Cf. \cite{Tihany} for
more details, especially in what concerns the hadronization model CAS
used in calculating results in Fig. 1.}. To better understand our
reasoning let us remind that classical weight method \cite{WE}
amounts to multiplying each event by special weight, i.e., event is
counted many times if it already possesses, by shear chance, traces
of desired bunching. In terms of philosophy of \cite{OMT}, which we
shall follow, it could be seen as selecting events already possesing
(to some degree, at least) a more bosonic character than other events
(i.e., in which particles are already bunched in way resembling that
of \cite{OMT}). What we propose is similar approach but performed
already on a single event level. Namely we propose to search for the
bosonic configurations of particles existing already in each event
because of the internal nonstatistical fluctuations provided by event
generator. Namely, there are groups of particles resembling those
obtained in \cite{OMT}, modulo only the fact that they usually have
different charges allocated to them whereas particles in \cite{OMT}
are of the same charge. We propose therefore to endow such bunches of
particles with the same charge to an extent limited only by the
overall charge conservation. This means that in cases where charge
allocation has been already provided by event generator we shall
neglect it and perform new charge allocation keeping, however, the
total number of particles of each charge the same as given by this
generator. Notice that we do it for each single event, keeping intact
both the original energy-momentum pattern provided by event generator
(i.e., conserve the energy-momentum) and all inclusive single
particle distributions. Leaving those interested in more details to
\cite{Tihany} we shall only say that to get desired result it is
enough to select one of the produced particles, allocate to it some
charge, and then allocate (in some prescribe way) the same charge to
as many particles located near it in the phase space as possible. In
this way one forms a cell in phase-space, which is occupied by
particles of the same charge only. This process should then be
repeated until all particles are used and it should be such that one
gets geometrical (Bose-Einstein) distribution of particles in a given
cell. This procedure changes the charge flow pattern provided by
event generator\footnote{It amounts to allowing formation of
multi-like-charged object on intermediate steps of hadronization
process. Therefore this method works only when such possibility
exists in a given generator.} retaining, however, both the initial
charge of the system and its total multiplicity distribution. The
procedure of formation of such cells is controlled by probability $P$
that given neighbor of the initially selected particle should be
counted as another member of the newly created emmiting cell in phase
space\footnote{It is important to realize that, because we do not
restrict {\it a priori} the number of particles which can be put in a
given cell, we are automatically getting BEC of {\it all orders}
(even if we use only two particle checking procedure at a given step
in our algorithm). It means that $C_2(Q=0)$ calculated in such
environment of the possible multiparticle BEC can exceed $2$.}. 

Referring to \cite{Tihany} for more details we shall illustrate
in Figs. 1a and 1b our attempts to describe (separately) $e^+e^-$
data by DELPHI on BEC \cite{Delphi} and intermittency \cite{DELPHI}
using the so called CAS model \cite{CAS} (see also \cite{Tihany})
whereas Figs. 1c and 1d show the respective intermittency and BEC
obtained when using parameters from the fits above. The results,
although not totally satisfactory, are encouraging given the
simplicity of CAS model used. Application of our method with other,
more sophisticated event generators should answer the question of its
final applicability. 

\section*{Acknowledgments}
Authors OU and GW would like to thank Wu Yuangfang and all Organizers
of XXXI-th ISMD for their kind hospitality and GW acknowledges
financial support received. The partial support of Polish Committee
for Scientific Research (grants 2P03B 011 18, 5P03B 091 21  and
621/E-78/SPUB/CERN/P-03/DZ4/99) is acknowledged.

\begin{figure}[t]
\epsfxsize=26pc 
\epsfbox{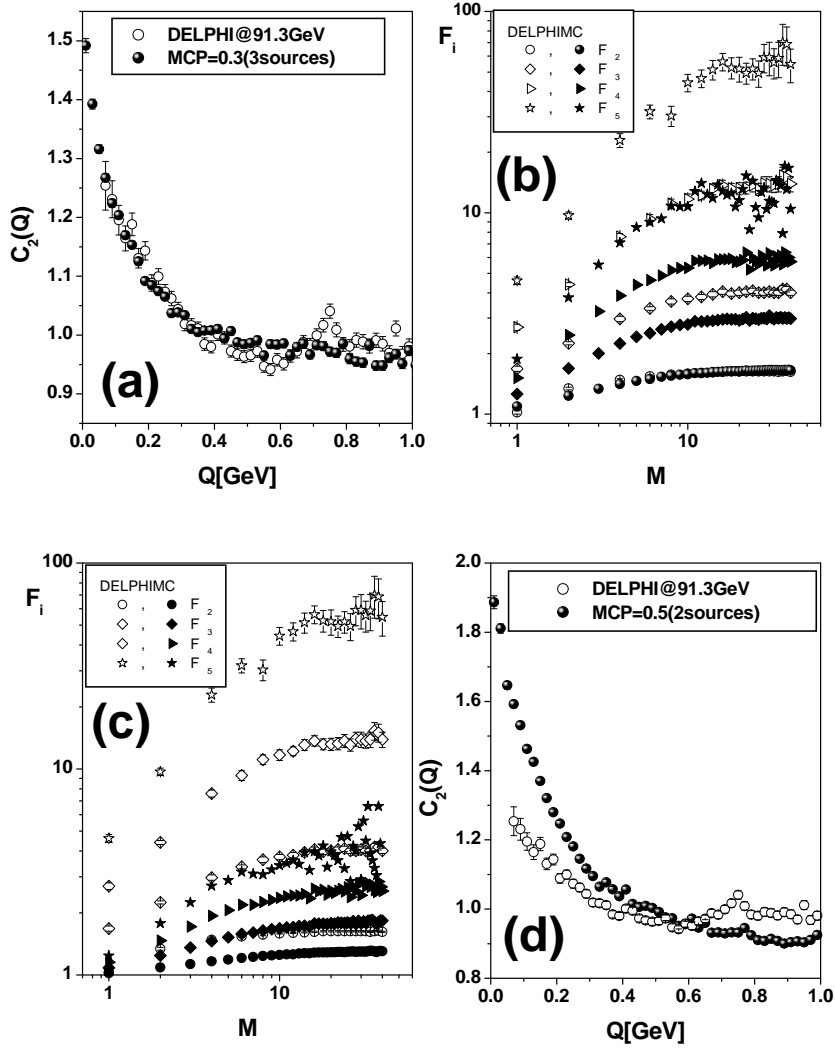} 
\vspace{-1cm}
\caption{The examples of best fits to data on $e^+e^-$ annihilation
by DELPHI on BEC $^8$ in $(a)$ and, separately, on factorial moments $^9$
$F_i$ in $(b)$ ($M$ is number of bins) using simple cascade
hadronization model CAS (cf. $^{7,10}$ for details). In $(c)$ are
shown factorial moments for parameters used in $(a)$ (when fitting
BEC) whereas in $(d)$ BEC for parameters used in $(b)$ (when fitting
factorial moments). Notice that left panels ($(a)$ and $(c)$) are
obtained for $3$ sources whereas right  panels ($(b)$ and $(d)$) for
$2$ sources. To fit $F_5$ one needs $P=0.75$ and two sources, but in
this case the calculated $F_2$ overshoots data by ca $50\%$.} 
\end{figure}


\begin{thebibliography}{99}

 \bibitem{BEC} R.M.Weiner, {\it Phys. Rep.} {\bf 327},
               249 (2000); U.A.Wiedemann and U.Heinz, {\it Phys.
               Rep.} {\bf 319}, 145 (1999);
               T.Cs\"org\H{o}, in {\it Particle Production Spanning
               MeV and TeV Energies}, eds. W.Kittel et al., NATO Science
               Series C, Vol. 554, Kluwer Acad. Pub. (2000), p. 203
               (see also: hep-ph/0001233).

 \bibitem{GEN} Cf., K.J.Escola, {\it On predictions of the first results
               from RHIC}, hep-ph/0104058, to be published in Proc.
               of QM2001, {\sl Nucl. Phys.} {\bf A} (2001).

 \bibitem{LS} L.L\"onnblad and T.Sj\"ostrand, {\it Eur. Phys. J.}
              {\bf C2} (1998) 165.

 \bibitem{WE} A.Bia\l as and A.Krzywicki, {\it Phys. Lett.} {\bf B354},
              (1995) 134; T.Wibig, {\it Phys. Rev.} {\bf D53} (1996) 3586;
              K.Fia\l kowski and R.Wit, {\it Eur. Phys. J.} {\bf C2},
              691 (1998) 691; B.Andersson and M.Ringn\'er, {\it Nucl.
              Phys.} {\bf B513} (1998) 627.

 \bibitem{AFTER} J.P.Sullivan et al., {\it Phys. Rev. Lett.} {\bf 70},
                 (1993) 3000; K.Geiger, J.Ellis, U.Heinz and
                 U.A.Wiedemann, {\it Phys. Rev.} {\bf D61} (2000) 
                 054002.

 \bibitem{OMT} T.Osada, M.Maruyama and F.Takagi, {\it Phys. Rev.}
               {\bf D59} (1999) 014024 (cf. also M.Biyajima,
               N.Suzuki, G.Wilk and Z.W\l o\-dar\-czyk, {\it Phys.
               Lett.} {\bf B386} (1996)) 297).

 \bibitem{Tihany} O.V.Utyuzh, G.Wilk and Z.W\l odarczyk, in the Proc.
                  of the XXX ISMD, Tihany, Hungary, 9-13 October 2000,
                  Eds. T.Cs\"org\H{o} et al., World Scientific 2001,
                  p. 373 (hep-ph/0101161, cf. also hep-ph/0102275).

 \bibitem{Delphi} P.Abreu et al. (DELPHI Collab.), {\sl Phys. Lett.} 
                  {\bf B286} (1992) 201. 

 \bibitem{DELPHI} P.Abreu et al. (DELPHI Collab.), {\sl Phys. Lett.} {\bf
                  B247} (1990) 137. 

 \bibitem{CAS} O.V.Utyuzh, G.Wilk and Z.W\l odarczyk, {\it Phys. Rev.}
               {\bf D61} (2000) 034007 and {\it Czech J. Phys.}
               {\bf 50/S2} (2000) 132 (hep-ph/9910355).

\end{thebibliography}
\end{document}